\documentclass[12pt,a4paper]{article}

\usepackage{amsmath}

\usepackage{amsfonts}
\usepackage{amssymb}

\title{Introduction to foundations of probability and randomness (for students in physics)\\
Lectures given at the Institute of  Quantum Optics and Quantum Information\\
Austrian Academy of Science\\ Lecture-1: Kolmogorov and von Mises}

\author{Andrei Khrennikov\\
International Center for Mathematical Modeling\\
 in Physics and Cognitive Sciences \\
 Linnaeus University,  V\"axj\"o-Kalmar, Sweden}

\begin{document}

\maketitle

\abstract{The education system for students in physics suffers (worldwide) from the absence of a deep course in probability and randomness.
This is the real problem for students interested in quantum information theory, quantum optics, and quantum foundations. Here the primitive
treatment of probability and randomness may lead to deep misunderstandings of theory and wrong interpretations of experimental results. Since 
during my visits (in 2013 and 2014) to IQOQI a number of students (experimenters!) asked me permanently about foundational problems of 
probability and randomness, especially inter-relation between classical and quantum structures, this year 
I gave two lectures on these problems. Surprisingly the interest of experiment-oriented students to mathematical peculiarities was very high.
This (as well as permanent reminding of prof. Zeilinger) motivated me to write a text based on these lectures which were originally presented 
in the traditional black-board  form. I hope that this might be useful for students from IQOQI as well as other young physicists.}

\section{Kolmogorov axiomatics of probability theory}

\subsection{Historical remarks}

We start with the remark that, opposite to geometry, axiomatic probability theory was created not so long ago. Soviet mathematician 
Andrei Nikolaevich Kolmogorov presented the modern axiomatics of probability theory only in 1933 in his book \cite{K}. The book was originally 
published in German.\footnote{The language and the publisher (Springer) were chosen by rather pragmatic reason. Springer paid by gold and 
young Kolmogorov needed money and gold was valuable even in Soviet Union.} The English translation \cite{K1} was published only in 1952 (and the Russian 
version \cite{K2} only in 1974). The absence of the English translation soon (when the German language lost its international dimension) led to the following problem.
The majority of the probabilistic community did not have a possibility to read Kolmogorov. Their picture of the Kolmogorov model was based
on its representations in English (and Russian) language textbooks. Unfortunately, in such 
representations a few basic ideas of Kolmogorov disappeared, since they were considered as just philosophic ``bla-bla'' having no value for mathematics. Partially this is correct, but
{\it probability theory is not just  mathematics.}  This is a physical theory and, as any physical theory, its mathematical formalism has to be endowed
with  some interpretation. In Kolmogorov's book the interpretation question was discussed in very detail. However, in majority of mathematical representations
of Kolmogorov's approach the interpretation issue is not enlighten at all.

It is interesting that the foundations of quantum probability theory were set practically at the same time as the foundations of 
classical probability 
theory. In 1935 John von Neumann \cite{VN} pointed that classical randomness is fundamentally different from quantum randomness. The first one 
is {\it ``reducible randomness''}, i.e., it can be reduced to variation of  features of systems in an ensemble. 
It can be also called {\it ensemble randomness.}
The second one is {\it 
``irreducible randomness''}, i.e., 
aforementioned ensemble reduction is impossible. By von Neumann quantum randomness is an {\it individual randomness}, even an individual electron 
 exhibits fundamentally random behavior. By him only quantum randomness is genuine randomness. 

Already now we emphasize that, although the notion of randomness is closely related to the notion of probability, they do not coincide.
We can say that the problems of creation of proper mathematical formalizations of classical and quantum probabilities were solved. However,
as we shall see, mathematics was unable (in spite of one hundred years of tremendous efforts) to provide proper formalization 
of neither classical nor quantum randomness. 

\subsection{The interpretation problem in quantum mechanics and classical probability theory} 

We turn to the interpretation problem of classical probability theory, see \cite{KHR_INT0}, \cite{KHR_INT} for a detailed presentation,  
in comparison with the interpretation 
problem of quantum mechanics (QM). The latter is well known, it is considered as one of foundational problems of QM. 
The present situation is characterized by huge diversity of interpretations. And this is is really unacceptable for a scientific theory. As all we know, those
interpretations are not just slight modifications of each other. They differ fundamentally: the Copenhagen interpretation of Bohr\footnote{It is not well known 
that originally Niels Bohr was convinced to proceed with the operational interpretation of 
QM by the Soviet physicist Vladimir Fock, the private communication 
of Andrei Grib who read the correspondence between Bohr and Fock in which Fock advertised 
actively the ``Copenhagen interpretation''.}, Heisenberg, Pauli (``shut up and calculate''),
the ensemble or statistical interpretation  of Einstein\footnote{Initially Schr\"odinger also kept to this interpretation. Nowadays 
it is practically forgotten that he elaborated his example with (Schr\"odinger) Cat and Poison just to demonstrate the absurdness of the Copenhagen 
interpretation. This example was, in fact, just a slight modification of Einstein's example with Man and Gun in his letter to Schr\"odinger. However, 
finally Schr\"odinger wrote to Einstein that he found very difficult if possible at all to explain the interference on the basis 
of the ensemble interpretation, so he gave up.}, Margenau, Ballentine (QM is a version of classical statistical mechanics with may be very tricky phase space), 
many worlds interpretation (no comment), ..., the  V\"axj\"o interpretation (the wave particle duality is resolved in favor of waves,
 a la early Schr\"odinger and the Devil is not in waves, but in detectors, a kind of combination of the 
ensemble and Copenhagen operational interpretations;
however,with ensembles of waves and not particles).\footnote{
We can point to the series of the V\"axj\"o conferences on quantum foundations where all possible interpretations were discussed and attacked 
during the last 15 years (see, e.g., lnu.se/qtap and lnu.se/qtpa for the last conferences and the book \cite{Beyond} devoted to this series). However, in spite of in general the great value 
of such foundational debates, the interpretation picture of QM did not become clearer.}   

Our main message is that {\it the problem of  finding of proper interpretations of classical probability and randomness is not less complex.} It is also characterized
by huge diversity of variants, e.g., measure-theoretic probability, frequency probability, subjective (Bayesian) probability, ... and randomness 
as unpredictability, randomness as complexity, randomness as typicality,.... 

It is interesting that nowadays in probability theory the problem of 
interpretation is practically ignored. One can say that the majority of the probability community proceed under the same slogan 
as the majority of the quantum community: ``shut up and calculate''.( Partially this is a consequence of the common 
treatment of probability theory as a theoretical formalism.
Questions of applicability of this formalism  are treated by statistics. Here the interpretation questions play an important role, may be even 
more important than in QM. By keeping to different interpretations (e.g., frequency and Bayesian)
researchers can come to very different conclusions based on the same statistical data; different interpretations generate different methods
of analysis of data, e.g., frequentists operate with confidence intervals estimates and Baeysians with credible intervals estimates. 
Opposite to QM, it seems that nowadays in probability theory and statistics nobody  expects that a ``really proper interpretation'' would be finally
created. 
  
The situation with randomness-interpretations is more similar to QM. It is characterized by 
practically one hundred years of discussions on possible interpretations of randomness. Different interpretations led 
to different theories of randomness. Numerous attempts to elaborate   a  ``really proper interpretation'' of randomness  
have not yet lead to a commonly acceptable conclusion. Nevertheless, there are still expectations that a new simple and unexpected
idea will come to life  and a rigorous mathematical model based on the commonly acceptable notion of randomness will be created
(this was one of the last messages of Andrei Kolmogorov before his death, the private communication of Albert Shiryaev).  

There is the opinion (not so common) that the interpretation problem of QM simply reflects the interpretation problem of 
probability/statistics/randomness; e.g., Zeilinger's {\it information interpretation} \cite{Z1} of QM (see also 
the series of works of Brukner and Zeilinger \cite{Z0}, \cite{Z1}, \cite{BR1}-\cite{BR3} and D' Ariano et al. \cite{D1}) is 
a quantum physical representation of Kolmogorov's complexity/information interpretation \cite{[134]a, [134], [135]}
 of classical randomness (also Chaitin \cite{Chaitin});
the Fuchs \cite{Fuchs, Fuchs1, Fuchs2} subjective probability interpretation of QM is nothing else than an application of the Bayesian interpretation of probability to QM, 
the V\"axj\"o interpretation \cite{Vaxjo1}, \cite{Vaxjo} 
is an attempt to apply to QM a combination of measure-theoretic and frequency interpretations of probability.

Finally, we remark that von Neumann considered the frequency interpretation probability of von Mises \cite{[169], [170], [171]} 
as the most adequate to the Copenhagen interpretation of QM
(a footnote in his book \cite{VN}, see also \cite{Beyond}).

\subsection{Events as sets and probability as measure on a family of sets representing events}

The crucial point of the Kolmogorov approach is the \textit{representation of random events by subsets} of some basic set $\Omega.$
This set is considered as {\it sample space} - the collection of all possible realizations 
of some experiment.\footnote{Consider an experiment corresponding to the $n$-times coin tossing. 
Each realization of this experiment generates a vector $\omega=\{x_1,...,x_n\},$ where $x_j=H$ or $T.$ Thus 
the sample space of this experiment contains $2^n$ points. We remark that this (commonly used)  sample space is based on outputs of 
the observable corresponding to coin's sides and not on so to say hidden parameters of the coin and the hand leading 
to these outputs. Later we shall discuss this problem in more detail.} 
Points of sample space are called {\it elementary events}.   

The collection of subsets representing random events
should be sufficiently rich -- to be able to perform the set-theoretic operations
such as the intersection, the union, and the difference of sets. However, at
the same time it should not be too extended. If a too extended system of
subsets is selected to represent events, then it may contain ``events'' which
cannot be interpreted in any  reasonable way.

After selection of a proper system of sets to represent events, one assigns
weights to these subsets:
\begin{equation}
\label{WEI}A \mapsto P(A).
\end{equation}
The probabilistic weights are chosen to be {\it nonnegative real numbers and
normalized by sum up to 1}: $P(\Omega)=1,$ the probability that something happens equals 
one. 

{\bf Interpretation:}  An event with large weight is more probable than an event with small
weight. (This is only a part of Kolmogorov's interpretation of probability, see section \ref{INTK}).

We now discuss another feature of the probabilistic weights:  the weight of an event $A$ that can be represented as the disjoint
union of events $A_{1}$ and $A_{2}$ is equal to the sum of weights of these
events. The latter property is called \textit{additivity}.%
\index{additivity}
(There is the evident similarity with mass, area, volume.)

It is useful to impose some restrictions on the system of sets representing
events: 

\begin{itemize}

\item a) set $\Omega$ containing all possible events and the empty set
$\emptyset$ are events (something happens and nothing happens);

\item b) the union of two sets representing events represents an event; 

\item c) the intersection of two sets representing events represents an event; 

\item d) the complement of a set
representing an event, i.e., the collection of all points that do not belong
to this set, again represents an event. 

\end{itemize}

These set-theoretic operations
correspond to the basic operations of classical {\it Boolean logics}: \textquotedblleft
or\textquotedblright, \textquotedblleft and\textquotedblright,
\textquotedblleft no\textquotedblright. And the modern set-theoretic
representation of events is a mapping of propositions describing events onto
sets with preservation of the logical structure.\footnote{At the beginning of the
mathematical formalization of probability theory the map (\ref{WEI}) was
defined on an algebraic structure corresponding to the logical structure, the
\textit{Boolean algebra} (invented by J. Boole, the creator of \textquotedblleft
Boolean logic\textquotedblright.)} The set-system with properties a)-d) is
called the \textit{algebra of sets} (in the American literature, the {\it field of
sets}).

\subsection{The role of countable-additivity ($\sigma$-additivity)}

In the case of finite $\Omega$ the map given by (\ref{WEI}) with the above-mentioned
properties gives the simplest example of  Kolmogorov's measure-theoretic probability.%
\index{probability}
(Since $\Omega$ can contain billions of points, this model is useful in a huge
class of applications.) Here $\Omega=\{\omega_{1},..., \omega_{N}\}.$ To
determine any map (\ref{WEI}), it is enough to assign to each point $\omega \in \Omega$
its weight
\[
0 \leq P(\omega_{j}) \leq1, \; \sum_{j} P(\omega_{j}) =1.
\]
Then by additivity this map is extended to the set-algebra consisting of all
subsets of $\Omega:$
\[
P(A) = \sum_{\{\omega_{j} \in A\}} P(\omega_{j}).
\]

\medskip

However, if $\Omega$ is countable, i.e., it is infinite and its points can
be enumerated, or ``continuous'' -- e.g., a segment of the real line
$\mathbf{R},$ then simple additivity is not sufficient to create a
fruitful mathematical model. The map (\ref{WEI}) has to be additive with
respect to countable unions of disjoint events ($\sigma$-additivity)\footnote{Here $\sigma$ is a symbol for
``countably''.}:
\begin{equation}
\label{LDIS}P( A_{1} \cup...\cup A_{n} \cup...)= P%
(A_{1})+...+ P(A_{n})+...,
\end{equation}
and to work fruitfully with such maps (e.g., to integrate), one has to impose
special restrictions on the system of sets representing events. 

It has to be
not simply a set-algebra, but a {\it $\sigma$-algebra of sets} (in the American
literature, a $\sigma$-field), i.e., b) and c) must be valid for countable unions
and intersections of sets. In the logical terms, it means that {\it the operations
``or'' and ``and'' can be applied infinitely many times to form new events.} 

Of course, this is a mathematical idealization of the real situation. 
Kolmogorov pointed out \cite{K, K1, K2} that, since in real experiments it is impossible to ``produce'' infinitely many
events, this condition is not experimentally testable. He would prefer to proceed with finite-additive probabilities.
However, without $\sigma$-additivity it is difficult (although possible)  to define the integral with respect a probability measure 
(Lebesgue integral is well defined only for $\sigma$-additive measures) and, hence, to define the mathematical expectation (the average
operation).

One of the most important ``continuous probability models'' is based on  the sample space $\Omega=
\mathbf{R},$ i.e.,  elementary events are represented by real numbers.
Typically a $\sigma$-algebra is selected as the \textit{Borel $\sigma
$-algebra:}%
\index{Borel sets}
it is generated by all half-open intervals, $[\alpha, \beta), \alpha< \beta,$
with the aid of the operations of the union, intersection, and complement.
We remark that in the case of ``continuous'' $\Omega$ it is not easy to define 
$\sigma$-algebras explicitly. Therefore one typically selects some simple system 
of subsets of $\Omega$ and then considers a minimal $\sigma$-algebra generated 
by this system.  

\subsection{Probability space}

Let $\Omega$ be a set and let $\mathcal{F}$ be a $\sigma$-algebra of its
subsets. A \textit{probability measure} $P $ is a map from
$\mathcal{F}$ to the segment $[0,1]$ normalized $P(
\Omega)=1$ and $\sigma$-additive, i.e., the equality (\ref{LDIS}) holds for
disjoint sets belonging to $\cal{F}.$

By the Kolmogorov axiomatics%
\index{Kolmogorov axiomatics}
\cite{K} the \textit{probability space}%
\index{probability space}
is a triple $\mathcal{P}=$
$(\Omega,$ $\mathcal{F},$ $P).$

Points $\omega$ of $\Omega$ are said to be \textit{elementary events},
elements of $\mathcal{F}$ are \textit{random events}, $P$ is
\textit{probability}.%
\index{probability}%

\medskip

{\bf Remark 1.} (Elementary events and random events). This terminology used by Kolmogorov
is a bit misleading. In fact, one has to distinguish elementary events from random events. 
The crucial point is that in general single point set $A_\omega= \{ \omega \},$ where $\omega$ is one 
of points of $\Omega,$ need not belong to the $\sigma$-algebra of events ${\cal F}.$ In such a case
we cannot assign the probability value to $A_\omega.$ Thus some elementary events are so to say hidden; 
although they are present mathematically at the set-theoretical level, we cannot assign to them the probability-values. One can consider
the presence of such hidden elementary events as a classical analog of hidden variables in QM, although 
the analogy is not complete.

\medskip

{\bf Example 1.} Let the sample space $\Omega=\{\omega_1, \omega_2,  \omega_3\}$ and let the collection (algebra) of events
${\cal F}$ consist of the following subsets of $\Omega: \{\emptyset, \Omega, A=\{\omega_1, \omega_2\}, B=\{\omega_3\}\},$ 
and $P(A)=P(B)=1/2.$ Here the 
elementary single point events $\{\omega_1\}, \{\omega_2\}$ are not ``physically approachable''. (Of course, this analogy 
is only indirect.) Interpretations of the existence of ``hidden elementary events'' were not discussed so much in 
classical probability. There exist realizations of the experiment (represented by the probability space ${\cal P}$ with this 
property) such that it is in principle impossible to assign probabilities to them, even zero probability. We shall come back to this 
problem by discussing the classical probabilistic representation of observables.

\section{Observables as random variables}

Random observations are represented by random variables. We start with
mathematically simplest random variables, the discrete ones. 

\textit{Discrete random variables}%
\index{random variable}
on the Kolmogorov space ${\cal P}$ are by definition functions $a:
\Omega\to X_{a} ,$ where $X_{a}=\{\alpha_{1},..., \alpha_{n},... \}$ is a
countable set (the \textit{range of values}) such that the sets
\begin{equation}
\label{WEIY}C_{\alpha}^{a}=\{\omega\in\Omega: a(\omega)= \alpha\}, \alpha\in
X_{a},
\end{equation}
belong to $\mathcal{F}.$ Thus classical probability theory is characterized by 
the {\it functional representation of observables.} For each elementary event $\omega \in \Omega,$ 
there is assigned the value of the observable $a,$ i.e., $\alpha= a(\omega).$  This is the $a$-observation on the realization $\omega$ 
of the experiment. 

It is typically assumed that the range of values $X_{a}$ is a subset of the
real line. We will proceed under this assumption.

\medskip

{\bf Remark 2.} (Observations and hidden elementary events). Let the system of events ${\cal F}$ do not contain the single-point set 
for some $\omega_0 \in \Omega.$ Thus the probability cannot be assigned to the $\omega_0$-outcome of the random experiment. 
Nevertheless,  this realization can happen and even the value of each observable is well defined: $a(\omega_0).$ However, we cannot 
``extract'' information about this elementary event with the class of observables corresponding to the selected probability space.
Consider the space from the Example 1; let $\omega \to a(\omega)$ be a random variable, $\alpha_j=a(\omega_j).$ Suppose that $\alpha_1
\not=\alpha_2.$ 
Then the sets $C_{\alpha_j}^{a}=\{\omega_j\}.$ However, the single point sets $\{\omega_j\}, j=1,2,$ do not belong to ${\cal F}.$ Hence,
we have to assume that $\alpha_1= \alpha_2.$ And this the general situation: we cannot distinguish such elementary events with the aid 
of observations.  

\medskip
 
The probability distribution%
\index{probability distribution}
of a (discrete) random variable $a$ is defined as $
p^{a}(\alpha) \equiv P(\omega\in\Omega: a(\omega)= \alpha).
$
The \textit{average}%
\index{average}
(mathematical expectation)%
\index{mathematical expectation}
of a random variable $a$ is defined as
\begin{equation}
\label{AVER0iii}\bar{a} \equiv E a=\alpha_{1} \; p^{a}(\alpha_{1}) +...+
\alpha_{n} \; p^{a}(\alpha_{n})+... .
\end{equation}
If the set of values of $\xi$ is infinite, then the average is well defined if
the series in the right-hand side of (\ref{AVER0iii}) converges absolutely.

\medskip

Suppose now that the results of random measurement cannot be represented by a
finite or countable set. Thus such an observable cannot be represented as a
discrete random variable.

A \textit{random variable}%
\index{random variable}
on the Kolmogorov space ${\cal P}$ is by definition any function $\xi:
\Omega\to\mathbf{R}$ such that for any set $\Gamma$ belonging to
the Borel $\sigma$-algebra, its pre-image belongs to the $\sigma$-algebra of
events $\mathcal{F}:$ $\xi^{-1}(\Gamma) \in\mathcal{F},$ where $\xi^{-1}(\Gamma)
=\{\omega\in\Omega: \xi(\omega) \in\Gamma\}.$

In this case the mathematical formalism is essentially more complicated. The
main mathematical difficulty is to define the integral with respect
to a probability measure on $\mathcal{F},$ the \textit{Lebesgue integral}
\cite{K}. In fact, classical probability theory, being based on
measure theory, is mathematically more complicated than quantum probability
theory. In the latter integration is replaced by the trace operation for
linear operators and the trace is always the discrete sum. Here we do not
plan to present the theory of Lebesgue integration. We formally invent the symbol
of integration.\footnote{We remark that, for a discrete random variable, the
integral coincides with the sum for the mathematical expectation, see
(\ref{AVER0iii}). And a discrete random variable is integrable if its
mathematical expectation is well defined. In general any integrable random
variable can be approximated by integrable discrete random variables and its
integral is defined as the limit of the integrals for the approximating
sequence.}

The \textit{average} (mathematical expectation) of a random variable $a$ is
defined as
\begin{equation}
\label{AVER0iii7}\bar{a} \equiv E a=\int_{\Omega}\xi(\omega) d P
(\omega).
\end{equation}
The probability distribution of a random variable $a$ is defined (for Borel
subsets of the real line) as $
p^{a}(\Gamma) \equiv P(\omega\in\Omega: a(\omega) \in\Gamma).$
This is a probability measure on the Borel $\sigma$-algebra. And the
calculation of the average can be reduced to integration with respect to
$
p^{a}: \bar{a} \equiv E a=\int_{\mathbf{R}} x \; d p^{a} (x).
$

\section{Conditional probability; independence}
\label{BS}

\textit{Kolmogorov's probability model} is based on a probability space
equipped with the operation of conditioning. In this model \textit{conditional
probability} is defined by the well known \textit{Bayes' formula}%
\index{Bayes' formula}
\begin{equation}
\label{BAE}P(B \vert C) = P(B\cap C)/P(C),
P(C)>0.
\end{equation}
By Kolmogorov's interpretation it is the \textit{probability of an event $B$
to occur under the condition that an event $C$ has occurred.} We emphasize that 
the Bayes formula is simply a definition, not a theorem; it is neither axiom
(the latter was stressed by Kolmogorov); we shall see that in the frequency approach 
to probability the Baeys formula is a theorem.  

We remark that conditional probability  (for the fixed conditioning $C))$ $P_{C}(B)\equiv P(B \vert C)$ is again a
probability measure on $\mathcal{F}.$ For a set $C\in\mathcal{F},
P(C)>0,$ and a (discrete) random variable $a,$ the conditional
probability distribution is defined as
$
p_{C}^{a}(\alpha) \equiv P(a= \alpha\vert C).
$
We naturally have
$
p_{C}^{a}(\alpha_{1})+...+ p_{C}^{a}(\alpha_{n})+... =1, \; p_{C}^{a}(\alpha_{n}) \geq 0.
$
The conditional expectation of a (discrete) random variable $a$ is defined by
$
E(a \vert C) = \alpha_{1} \; p_{C}^{a}(\alpha_{1}) + ...+\alpha_{n} \; p_{C}^{a}(\alpha_{n})+....
$
\medskip

Again by definition two events $A$ and $B$ are {\it independent} if 
\begin{equation}
\label{YYYTB7}
P(A\cap B)= P(A) P(B)
\end{equation}
In the case of {\it nonzero probabilities} $P(A), P(B)>0$ independence can be formulated in terms of 
conditional probability:
\begin{equation}
\label{YYYTB77}
P(A\vert B)= P(A) 
\end{equation}
or equivalently 
\begin{equation}
\label{YYYTB77a}
P(B\vert A)= P(B). 
\end{equation}
The {\it relation of independence is symmetric:}
if $A$ is independent of $B,$ i.e., (\ref{YYYTB77}) holds,  then $B$ is independent of $A,$ i.e., (\ref{YYYTB77a}) holds, and vice versa. 
(We remark that this property does not match completely with our intuitive picture of independence.)

\subsection{Formula of total probability}

\label{TYY}

In our further considerations the important role will be played by the
\textit{formula of total probability}%
\index{formula of total probability}
(FTP). This is a theorem of the Kolmogorov model. Let us consider a countable
family of disjoint sets $A_{k}$ belonging to $\mathcal{F}$ such that their
union is equal to $\Omega$ and $P(A_{k}) >0, k=1,....$ Such a family
is called a \textit{partition}%
\index{partition}
of the space $\Omega.$

\medskip

\textbf{Theorem 1.} \textit{Let $\{A_{k} \}$ be a partition. Then, for every
set $B \in\mathcal{F},$ the following formula of total probability holds}
\begin{equation}
\label{TP}P(B)= P(A_{1}) P(B \vert A_{1}) +...+
P(A_{k}) P(B \vert A_{k})+....
\end{equation}

\textbf{Proof.} We have
\[
P(B)=P(B\cap\left(  \cup_{k=1}^{\infty}A_{k}\right)
)=\sum_{k=1}^{\infty}P(B\cap A_{k})=\sum_{k=1}^{\infty}%
P(A_{k})\frac{P(B\cap A_{k})}{P(A_{k})}.
\]

\medskip

Especially interesting for us is the case where a partition is induced by
a discrete random variable $a$ taking values $\{\alpha_{k}\}.$ Here,
\begin{equation}
\label{RV1}A_{k}= \{\omega\in\Omega: a(\omega)= \alpha_k\}.
\end{equation}
Let $b$ be another random variable. It takes values $\{\beta_{j}\}.$ For any
$\beta\in X_{b},$ we have
\begin{equation}
\label{TP1}
P(b= \beta)= \sum_k P(a=\alpha_{k}) P(b= \beta\vert a=\alpha_{k})
\end{equation}

This is the basic formula of the Bayesian analysis. The value of  the probability 
to obtain the result $\beta$ in the $b$-measurement can be estimated on the basis 
of conditional probabilities for this result (with respect to the results of the $a$-measurement).

\subsection{The Kolmogorov strong law of large numbers}

Consider a sequence of identically distributed and independent random variables
 $\xi_1,...,\xi_N,...,$ with average $m=E \xi_j.$ Then by the {\it strong law of large numbers} we have: 
\begin{equation}
\label{LLN0}
P\Big(\omega\in \Omega : \frac{\xi_1(\omega) +...+\xi_N(\omega)}{N} \to m, N \to \infty \Big)=1,
\end{equation}
i.e., 
$$
\lim_{N\to \infty} \frac{\xi_1(\omega) +...+\xi_N(\omega)}{N} = m
$$
almost everywhere (on the set of probability 1). 

This law is the basic law guaranteeing the applicability of the Kolmogorov measure-theoretic model
to experimental data. It was proven by Kolmogorov and this was the important step in justification 
of his model.

The main interpretation problem related to this law is of  the following nature. 
Although the Kolmogorov strong law of large numbers implies that the set of elementary events for which the arithmetic mean converges
to the probabilistic mean $m$ is very large (from the measure-theoretic viewpoint), for the concrete 
$\omega$ (concrete sequence of experimental trials), this law cannot provide any information whether this convergence take place or not.
In fact, this was the main argument of von Mises (the creator of frequency probability theory) against the Kolmogorov measure-theoretic 
model, see the remark at the very end of section \ref{VMISES}.

Consider now some event $A,$ i.e., $A \in \mathcal{F},$ and generate a sequence of independent tests in which the event $A$ either happen 
or not. In the Kolmogorov model this sequence of experimental trials is described by a sequence of independent random variables,
$\xi_j =1$ if $A$ happens and $\xi_j =0$ in the opposite case. Then the frequency of the $A$-occurence, $\nu_N(A)=n(A)/N,$ can be represented as 
$$
\nu_N(A) =     \frac{\xi_1(\omega) +...+\xi_N(\omega)}{N}.
$$
Hence, the strong law of large numbers, in particular, implies that probabilities of events are approximated by relative frequencies.

\subsection{Kolmogorov's interpretation of probability}
\label{INTK}

Kolmogorov  proposed \cite{K} to interpret probability as follows: ``[. . . ] we may assume
that to an event $A$ which may or may not occur under conditions $\Sigma$ is assigned a real
number $P(A)$ which has the following characteristics: 
\begin{itemize} 

\item (a) one can be practically certain
that if the complex of conditions $\Sigma$ is repeated a large number of times, $N,$ then
if $n$ be the number of occurrences of event $A,$ the ratio $n/N$ will differ very slightly
from $P(A);$

\item (b) if $P(A)$ is very small, one can be practically certain that when conditions
$\Sigma$ are realized only once the event $A$ would not occur at all.'' 

\end{itemize}

The (a)-part of this interpretation is nothing else than the frequency interpretation of probability, cf. with von Mises
theory and his principle of the statistical stabilization of relative frequencies, section \ref{VMISES}. In the measure-theoretic 
approach this viewpoint to probability is justified by the {\it law of large numbers.} However, for Kolmogorov approximation 
of probability by frequencies was not the only characteristic feature of probability. The (b)-part also plays an important role.
This is the purely weight-type argument: if the weight assigned to an event is very small than one can expect that such an event 
will practically never happen. (Of course, here the meaning of ``practically'' is the subject for a discussion.)
 
\subsection{Random vectors; existence of joint probability distribution}    
 
 In the Kolmogorov model a system of observables is represented by a vector of 
random variables $a=(a_1,..., a_n).$ Its probability distribution is defined as 
\begin{equation}
\label{PD7}
p_{a}(A_1\times \cdots \times A_n) = P(\omega: a_1(\omega) \in A_1,\cdots, a_n(\omega) \in A_n).
\end{equation}
Thus, for any finite system of observables, it is assumed that the joint probability distribution 
exists. 

We remark that, for any subset of indexes $i_1,...,i_k,$ the probability distribution of the vector 
$(a_{i_1},...,a_{i_k}),$ $p_{a_{i_1},..., a_{i_k}}(A_{i_1}\times...\times A_{i_k}) = 
P(\omega \in \Omega: a_{i_1}(\omega) \in A_{i_1}, ..., a_{i_k}(\omega)  \in A_{i_k}),$  can be obtained 
from $p_a$ as its marginal distribution. For example, consider a pair of discrete observables  
$a=(a_1,a_2).$ Then
\begin{equation}
\label{LLL}
p_{a_{1}}(\alpha_1) = \sum_{\alpha_2} p_a(\alpha_1, \alpha_2); \;
p_{a_{2}}(\alpha_2) = \sum_{\alpha_1} p_a(\alpha_1, \alpha_2).
\end{equation} 
In discussions related to Bell's inequality these conditions of the marginal consistency are known
as {\it no-signaling conditions.}

Consider now a triple of discrete observables  $a=(a_1,a_2, a_3).$ Then
\begin{equation}
\label{LLL1}
p_{a_{1}, a_{2}}(\alpha_1, \alpha_2) = \sum_{\alpha_3} p_a(\alpha_1, \alpha_2, \alpha_3),..., 
p_{a_{2}, a_{3}}(\alpha_2, \alpha_3) = \sum_{\alpha_1} p_a(\alpha_1, \alpha_2, \alpha_3).
\end{equation} 
One dimensional distributions can be obtained from  the two dimensional distributions in the same way 
as in (\ref{LLL}). 

We remark that the possibility to represent the two dimensional probability distributions as marginals of the joint probability 
distribution is the basic assumption for derivation of the Bell-type inequalities. In the second lecture 
we shall discuss this point in more detail.

We remark that in probability theory the problem of representation of a family of observables by 
random variables on the same probability space has the long history. Already G. Boole \cite{Boole}, \cite{Boole1} considered 
the following problem. There are given three observables $a_1, a_2, a_3=\pm 1$ and their pairwise 
probability distributions $p_{a_i a_j}.$ Is it possible to find a discrete probability measure $p(\alpha_1, \alpha_2, 
\alpha_3)$ such that all $p_{a_i a_j}$ can be obtained as the marginal probability distributions? And he showed 
that the inequality which nowadays is known as the Bell inequality is a {\it necessary  condition} for the 
existence of such $p.$ (At that time measure theory had not yet been well developed and he used the algebraic formalism for probability
based on Boolean algebras.) For Boole, it was not self-evident that any family of observables can be described with the aid
of a single probability space. He understood well that if data can be collected only for pairwise measurements, but 
it is impossible to measure jointly the triple, then it can happen that the ``Boole-Bell inequality'' is violated. For him,
such a violation does not mean a violation of the laws of classical probability theory. From the very beginning 
classical probability theory was developed as theory about {\it observable events} and the corresponding probabilities.
Roughly speaking probability distributions of ``hidden variables'' and counterfactuals were completely foreign for the 
creators of the probability theory.   A. N. Kolmogorov pointed out \cite{K1}, section 2, that each complex of experimental 
physical conditions determines its own probability space. In the general setup he did not discuss the possibility to represent the data 
collected for different (may be incompatible) experimental conditions with the aid of a single probability space, cf. with Boole \cite{Boole},
\cite{Boole1}. Therefore we do not know his opinion about this problem. However, he solved positively this problem in one very important case.
This is the famous Kolmogorov theorem guaranteeing the existence of the probability distribution for any stochastic process, see section
\ref{STPR}.

In 1962 the Soviet probabilist  N. N. Vorob'ev \cite{V} presented the complete solution of the problem of the existence of 
the joint probability distribution for any family of discrete random variables.  He used his criteria for problems of game theory 
(the existence of a mixed strategy) and in random optimization problems (the existence of an optimal solution). However, the reaction of 
the Soviet probabilistic school (at that time it was one of the strongest in the world) to his attempt 
to to go beyond the Kolmogorov axiomatics was very negative. His work \cite{V} was completely forgotten. Only in 2005 it was ``found''
by Karl Hess and Walter  Philipp and used in the Bell debate.   We remark that in the probabilistic community  the general tendency was 
to try to find conditions of the existence a single probability space. Therefore the works in which non-existence was the main point 
were (consciously or unconsciously) ignored. In particular, even the aforementioned works of G. Boole were also completely forgotten.
Around 2000 they were ``found'' by Itamar Pitowsky who used them in the Bell debate.   

Finally, we formulate the following fundamental problem: {\it Can quantum probabilistic data be described by the classical (Kolmogorov)
probability model?} 

It seems that, since the (Boole-)Bell inequality is a necessary condition for the existence of such a description and since it is violated for 
the quantum probabilistic data, these data cannot be embedded in the Kolmogorov model. However, the real situation is more complicated
and it will be discussed in the second lecture, see also \cite{KHR_ARXIV}.  

\subsection{Stochastic processes} 
\label{STPR}

The notion of a random vector is generalized to the notion of a {\it stochastic process.} 
Suppose that the set of indexes is infinite; for example,  $a_t, t \in [0,+\infty).$ Suppose that, for 
each finite set $(t_1...t_k),$ the vector $(a_{t_1} ... a_{t_k})$ can be observed and its probability 
distribution $p_{t_1...t_k}$ is given. By selecting $\Omega_{t_1...t_k}= \mathbf{R}^k,$ 
$P_{t_1...t_k}= p_{t_1...t_k},$ and ${\cal F}_k$ as the Borel $\sigma$-algebra for $\mathbf{R}^k,$
 we obtain the probability space ${\cal P}_{t_1...t_k}$
describing measurements at points $t_1...t_k.$   
At the beginning of 20th century
the main mathematical  question of probability theory was whether it is possible to find a single probability space 
${\cal P} =(\Omega, {\cal F}, P)$ such that all $a_t$ be represented as random variables on this space and all probability 
distributions $p_{t_1...t_k}$ are induced by the same $P:$ 
$$
p_{t_1...t_k}(A_1\times \cdots \times A_{k}) =P(\omega \in \Omega: a_{t_1}(\omega) \in A_{1},  , a_{t_n}(\omega)  
\in A_{n}).
$$ 
Kolmogorov found natural conditions for the system of measures $p_{t_1...t_k}$ which guarantee existence 
of such a probability space, see \cite{K}.\footnote{We remark that 
the $\Omega$ is selected as the set of all trajectories $t \to \omega(t).$ 
The random variable $a_t$ is defined as $a_t(\omega) =\omega (t).$
Construction of the probability
measure $P$ serving for all finite random vectors is mathematically advanced and going back to construction of the Wiener 
measure on the space of continuous functions.}  
And during the next 80 years analysis of properties of (finite and infinite) families of random variables defined on one fixed  probability 
space was the main activity in probability theory.

The Kolmogorov conditions are necessary and sufficient and their are formulated as
\begin{itemize}

\item For any permutation $s_1...s_k$ of indexes $t_1...t_k,$
$$
p_{t_1...t_k} (A_{t_1} \times...\times A_{t_k}) = p_{s_1...s_k} (A_{s_1} \times...\times A_{s_k}) .
$$
\item For two sets of indexes $t_1...t_k$ and $r_1,..,r_m,$
$$
p_{t_1...t_k r_1...r_m} (A_{t_1} \times...\times A_{t_k} \times \mathbf{R}\times... \times  \mathbf{R})= 
p_{t_1...t_k} (A_{t_1} \times...\times A_{t_k}).
$$
\end{itemize}

\section{Frequency (von Mises) theory of probability}
\label{VMISES}

Von Mises (1919) theory was the first probability theory \cite{[169], [170], [171]} based fundamentally on 
the principle of the stabilization of statistical frequencies. Although this principle
was heuristically used from the very beginning of probabilistic studies, only von Mises
tried to formalize it mathematically and to put it as one of the basic principles 
of probability theory. His theory is based on the notion of a
{\it collective} ({\it random sequence}). 

Consider a random experiment $S$ and denote by $L=\{\alpha_1,..., \alpha_m\}$ the set
of all possible results of this experiment\footnote{R. von Mises did not consider probability theory as 
a purely mathematical theory. He emphasized that this is a physical theory such as, e.g., hydrodynamics. 
Therefore his starting point is a physical experiment which is a structure from physics and not from mathematics.
He was criticized for mixing physics and mathematics. But he answered that there is no way to proceed with probability 
as a purely mathematical entity, cf. with remark of A. Zeilinger on the notion of randomness, section \ref{FINAL}.}. 
The set $L$ is said to be the label set, or the set
of attributes of the experiment $S.$ We consider only finite sets $L.$ Let us consider $N$ trials  for this 
$S$ and record its results,  $x_j \in L.$  This process generates a finite sample:
\begin{equation}
\label{3.1}
x=(x_1,..., x_N\},  x_j \in L.
\end{equation}
A {\it collective} is an infinite idealization of this finite sample:
\begin{equation}
\label{3.2}
x=(x_1,..., x_N,...\},  x_j \in L,
\end{equation}
for which the following two von Mises' principles are valid.

\medskip

${\bf Principle \; 1}$ (statistical stabilization).

This is the principle of the statistical stabilization of relative frequencies of each attribute $\alpha \in L$ 
of the experiment $S$ in the sequence (\ref{3.2}). Take the frequencies 
$$
\nu_N(\alpha;x) = \frac{n_N(\alpha;x)}{N} 
$$
where $\nu_N(\alpha;x)$ is the number of appearance of the attribute $\alpha$ in the first $N$ trials. By the
principle of the statistical stabilization
 {\it the frequency $\nu_N(\alpha;x)$ approaches a limit as $N$ approaches infinity for every label $\alpha\in L.$} 

This limit 
$$
P_x(\alpha)= \lim _{N \to \infty} \nu_N(\alpha;x)
$$
is called {\it the probability of the attribute $\alpha$ 
of the random experiment $S.$}

Sometimes (when the random experiment and the collective generated by it are fixed) 
this probability will be denoted simply as $P(\alpha)$ (although von Mises would be really angry; he always
wrote that operation with abstract probabilistic symbols, i.e., having no relation to random experiment, 
is meaningless and may lead to paradoxic conclusions). We now cite von Mises \cite{[170]}:

{\small ``We will say that a collective is a mass phenomenon or a repetitive event, or simply
a long sequence of observations for which there are sufficient reasons to believe
that the relative frequency of the observed attribute would tend to a fixed limit if the
observations were infinitely continued. This limit will be called the probability of the
attribute considered within the given collective''.}

\medskip

${\bf Principle \; 2}$ (randomness).

Heuristically it is evident that we cannot consider, for example, the sequence 
$$
x=(0, 1, 0, 1, 0, 1,...)
$$
as the output of a random  experiment. However, the principle of
the statistical stabilization holds for $x$ and $P_x(0)=P_x(1)=1/2.$ 
To consider  sequences (\ref{3.2}) as objects of probability theory,  
we have to put an additional constraint on them:

{\it The limits of relative frequencies have to be stable with respect to a place selection (a
choice of a subsequence) in (\ref{3.2}).}

\medskip

In particular, $z$ does not satisfy this principle.\footnote{A. Zeilinger commented that, although this sequence 
has the deterministic structure, in principle it can be generated by some intrinsically random experiment $S.$ May be 
the probability of such output is zero, but the experimenter would be ready to see such an event. This statement matches 
well with the ideology of the Kolmogorov measure-theoretic approach, but not of the von Mises frequency approach. By the latter
if one obtained such a sequence, she has to question randomness of her experiment.}

However, this very natural notion (randomness) was the hidden bomb in the foundations of von
Mises' theory. The main problem was to define a class of place selections which
induces a fruitful theory. The main and very natural restriction which was set by von Mises is that a place selection
in (\ref{3.2}) cannot be based on the use of attributes of elements. For example, one cannot
consider a subsequence of (\ref{3.2}) constructed by choosing elements with the fixed label
$\alpha \in L.$  Von Mises  defined a place selection in the following way \cite{[170]}, p.9:

\medskip

${\bf PS}$ {\small ``a subsequence has been derived by a place selection if the decision to retain
or reject the $n$th element of the original sequence depends on the number $n$ and
on label values $x_1,..., x_{n-1}$  of the $n-1$ preceding elements, and not on the
label value of the $n$th element or any following element''.}

\medskip

Thus a place selection can be defined by a set of functions 
$$
f_1, f_2(x_1), f_3(x1, x2), f_4(x_1, x_2, x3),...
$$
each function yielding the values 0 (rejecting the $n$th
element) or 1 (retaining the $n$th element).

Here are some examples of place selections: 
\begin{itemize}

\item choose those $x_n$ for which $n$ is
prime; 

\item choose those $x_n$ which follow the word $01;$ 

\item toss a (different) coin;
choose $x_n$ if the $n$th toss yields heads. 

\end{itemize}

The first two selection procedures are law-like, the third selection random. 
It is clear that all of these procedures
are place selections: the value of $x_n$ is not used in determining whether to choose $x_n.$

The principle of randomness ensures that no strategy using a place selection rule
can select a subsequence that allows different odds for gambling than a sequence that
is selected by flipping a fair coin. Hence, can be called {\it the law of excluded
gambling strategy.}

\medskip

{\bf Remark 3.} The Kolmogorov measure-theoretic model is solely about probability; 
there is nothing about randomness. In the von Mises frequency model randomness and probability 
cannot be separated.   

\medskip

{\bf Remark 3.} Later we shall present a variety of approaches to the notion of randomness. The von Mises
approach can be characterized as {\it randomness as unpredictibility} (no chance to beat the roulette).

\medskip

The definition (${\bf PS}$) suffered from the absence of the mathematical proof of the existence of collectives.
Von Mises reaction to the critique  from mathematicians side was not constructive (from the mathematicians viewpoint). He told:
go to casino and you will get a plenty of random sequences, cf. with Zeilinger's proposal on physical randomness section \ref{FINAL}. 

\medskip

One can summarize the result of extended mathematical studies on the von Mises notion of randomness as follows:

{\it If a class of place selections is too extended then the notion of the collective is too restricted (in fact, there
are no sequences where probabilities are invariant with respect to all place selections).
Nevertheless, by considering invariance of probabilities with respect to only special classes of place selections
it is possible to construct sufficiently reach set of such ``restricted collectives''.}

And  von Mises himself was completely satisfied with the latter operational solution of
this problem. He proposed \cite{[171]} to fix a class of place selections
which depends on the physical problem and consider  the sequences of attributes in which
probabilities are invariant with respect of only this class of selections.  Thus he again tried to remove
this problem outside the mathematical framework.

\medskip

The frequency theory of probability is not, in fact, the calculus of probabilities, but
it is the calculus of collectives which generates the corresponding calculus of probabilities.
We briefly discuss some of the basic operations for collectives, see \cite{[171]}, \cite{KHR_INT} for
the detailed presentation. We remark from the very beginning that operations for collectives are more complicated
than the set theoretical operations. We consider only one operation.   

\medskip

{\it Operation of mixing of collectives and the basic properties of probability.}
 Let $x$ be a collective with the (finite) label space $L_x$ (here it is convenient to index label spaces by collectives) and
let $E=\{\alpha_{j1},...,\alpha_{jk} \}$  be a subset of $L_x.$ The sequence (\ref{3.2})  is transformed into
a new sequence $x_E$ by the following rule (this is the operation called by von Mises {\it mixing}). 
If $x_j \in E,$ then at the $j$th place there is written 1; if $x_j\not\in E$ then
there is written  0. Thus the label set of $x_E$ consists of two points, $L_{x_E}=\{0,1\}.$ This sequence has
the property of the statistical stabilization for its labels. For example,
$$
P_{x_E}(1)= \lim_{N \to \infty} 
\nu_N(1;x_E) =\lim_{N \to \infty} 
\nu_N(E;x)= \lim \sum_{n=1}^k \nu_N(\alpha_{jn};x),
$$
where $\nu_N(E;x)=\nu_N(1;x_E)$ is the relative frequency of $1$ in $x_E$ which coincides 
with the relative frequency of appearance of one of the labels from $E$ in the original collective 
 $x.$
Thus
\begin{equation}
\label{3.3}
P_{x_E}(1)=\sum_{n=1}^k P_x(\alpha_{jn}).
\end{equation}
 To obtain (\ref{3.3}), we have only used the fact that {\it the addition is a continuous operation
on the field of real numbers}\footnote{At the moment this remark that properties of (frequency) probability have some coupling with 
inter-relation of the algebraic and topological structures on the real line can be considered simply as a trivial  statement of  
 purely mathematical nature.
However, later we shall see that the result that {\it additivity of probability is a consequence of the fact that $\mathbf{R}$ is an additive 
topological group} has a deep probabilistic meaning and it can lead to creation of a family of generalized frequency probability models 
a la von Mises, section \ref{APP}.}
  ${\bf R}.$ 

It is possible to show that the sequence $x_E$ also satisfies the
principle of randomness.\footnote{Here we do not discuss the problem of existence
of collectives. Suppose that they exist or consider some special class of place selections --  the restricted principle of randomness, 
see section \ref{RRR}.} 
Hence this is a new collective. By this operation
any collective $x$ generates a probability distribution on the algebra ${\cal F}_{L_x}$ of all subsets
of $L_x$ (we state again that we consider only finite sets of attributes): 
\begin{equation}
\label{3.3a}
P(E) (\equiv P_x(E)) = P_{x_E} (1)=\sum_{n=1}^k P_x(\alpha_{jn}).
\end{equation}
We present the properties of this probability. 

\begin{itemize}

\item ${\bf P1}$ {\it Probability takes values in the segment $[0,1]$ of the real line.}

Since $P(E) = \lim_{N\to \infty} \nu_N(E; x)= \lim_{N\to \infty} \frac{n_N(E; x)}{N},$ 
and $0 \leq \nu_N(E; x) = \frac{n_N(E; x)}{N} \leq 1,$ then $ P(E) \in [0,1].$

\item ${\bf P2}$ {\it Probability that something happens equals to 1.}

This is a consequence of the evident fact that for $E= L_x$ the collective $x_E$ does not contain zeros.

\item ${\bf P3}$ {\it Additivity of probability.}

Here we use the representation (\ref{3.3a}) of probability. If $E_1$ and $E_2$ are two disjoint subsets 
of $L_x,$ then, for $E=E_1\cup E_2,$ 
$$
P(E)= \sum_{\alpha\in E} P(\alpha) = \sum_{\alpha\in E_1} P(\alpha) + \sum_{\alpha\in E_2} P(\alpha)= 
P(E_1) + P(E_2).
$$
\end{itemize}

Thus in the von Mises theory probability is also a probability measure. However, opposite 
to the Kolmogorov theory, properties ${\bf P1} - {\bf P3}$ are {\it theorems}, not axioms.
It is interesting that in his book \cite{K} Kolmogorov referred to von Mises theory to justify 
the properties ${\bf P1} - {\bf P3}$ for the probability measure.  

\medskip

We recall that in the Kolmogorov model the Bayesian expression of conditional probability is simply 
the definition. In the von Mises model this is again a theorem (based on another operation for collectives, so called 
partitions of collectives, see  \cite{[169], [170], [171]}). Thus heuristically the frequency theory is better justified. However, formally
it is less rigorous (see \cite{KHR_INT} for an attempt to lift this theory to the mathematical level of rigorousness).

We can say that the von Mises model was {\it the first operational model of probability.} In some sense it has even higher 
level of operationalization than the quantum model of probability. In the latter randomnesses carried by a system and a measurement device are still
separated, one is represented by a quantum state (density operator) and another by an observer (Hermitian operator or POVM). Von Mises collective
unifies these randomnesses.

Since the frequency definition of probability induces the probability measure on the algebra of all subsets of the set 
$L_x$ of the attributes of a random experiment inducing the collective $x,$ one may think that the Kolmogorov model can be 
considered simply  as emergent from the von Mises model. And von Mises advertised  this viewpoint in his books. Even the Kolmogorov
referring to the von Mises model to justify the axioms of the measure-theoretic probability may induce such an impression. However,
this is  not so simple issue. Kolmogorov's sample space $\Omega$ (endowed with the corresponding ($\sigma$)-algebra of random events)
is not the same as von Mises' space of experiments attributes. For repeatable experiments, 
Kolmogorov's elementary events are, in fact, sequences of trials, so roughly speaking Kolmogorov's probability is defined 
on the set of all possible von Mises collectives. Of course, since the Kolmogorov model is abstract, one can consider even 
probabilities on the set of experiment's attributes, as von Mises did. However, the latter would not correspond to repeatable 
experiments in the Kolmogorov scheme. 

The von Mises model can be easily extended to the case of countable sets of 
the experiment attributes, $L=(\alpha_1,...., \alpha_n,....).$
However, extension to ``continuous sets of attributes'', e.g., $L={\bf R},$ 
is mathematically very difficult. It will not lead 
us the the notion of a measurable set used in the theory of the Lebesgue integral. 

In the measure-theoretic  model of Kolmogorov the (strong) law of large numbers is interpreted as providing the 
frequency interpretation of probability. Von Mises did not agree with such an interpretation. He claimed that, opposite 
to his principle of the statistical stabilization guarantying convergence of relative frequencies to the probability 
in each concrete collective (random sequence), the  law of large numbers is a purely mathematical statement which has
no direct relation to experiment. His main critical argument was that knowing something with probability one is totally
meaningless if one need to know something about the concrete random sequence.

\medskip

Thus one can say that {\it in  the development of  probability we selected the pathway based on a simpler mathematics and rejected the pathway 
based on our heuristic picture of probability and randomness.}

\section{Random sequences}

\subsection{Approach of Richard von Mises: randomness as unpredictability}
\label{RRR}

Von Mises did not solve the problem of the existence of collectives. Immediately after his proposal on collectives, mathematicians started 
to ask whether such sequences exist at all. One of the first objection was presented by E. Kamke, see, e.g.,  \cite{KHR_INT} and it is well known as  
{\it Kamke's objection.}

The principle of randomness of von Mises implies on stability of
limits of frequencies with respect to the set of {\it all possible place selections.}
Kamke claimed that there are no sequences  satisfying this principle. Here we reproduce 
his argument.

\medskip

Let $L=\{0,1\}$ be the label set and let $x=(x_1,...,x_n,...), x_j \in L,$ 
be a collective which induces the probability distribution $P: P(0)=P(1)=1/2.$
Now we consider the set $S_{\rm{increasing}}$ of all strictly
increasing sequences of natural numbers, its elements have the form 
$k=(n_1 < n_2 < n_m <...),$ where $n_j,j=1,2,...$ are natural numbers. 
This set can be formed independently of the collective $x.$ However,
among elements of $S_{\rm{increasing}}$, we can find the strictly increasing sequence $\{n: x_n=1\}.$  This
sequence define a place selection which selects the subsequence $(1,1,...,1,...)$  from the sequence $x.$
Hence, the sequence $x$ (for which we originally supposed that it is a collective) is not a collective after all!

\medskip

However, the mathematical structure of Kamke's argument was not completely convincing.   
He claimed to have shown that for every putative collection $x$ there exists a
place selection $\phi$ that changes the limits of frequencies.
But the use of the existential quantifier here is classical (Platonistic). Indeed, it seems
impossible to exhibit explicitly a procedure which satisfies von Mises' criterion (independence
on the value $x_n)$ and at the same time selects the subsequence $(1,1,1,...)$ from the original sequence
$x.$ In any event Kamke's argument played an important role in understanding that to create 
the mathematically rigorous theory of collectives one has to restrict the class of place selections.\footnote{The work of Kamke as well
as other works devoted to the analysis of the von Mises principle of the place selection also played the fundamental role in setting the foundations
of theory of algorithms and constructive mathematics. In the latter the arguments similar to the Kamke objection are taken into account,
because they are not based on constructive (algorithmically performable) proofs.}

The simplest way (at least from the mathematical viewpoint) is to  proceed with special classes of 
{\it lawlike place selections.} In particular, A. Church proposed to consider place selections
in which the selection functions $f_n(x_1,...,x_{n-1})$ are algorithmically computable. (We recall that $f_n$ is used to select/not the 
$n$th element of a sequence $x=(x_1,...,x_n,...).$ It is important to note that the set of all Church-like place selections
is countable, see, e.g., \cite{INT_KHR}.   The existence of 
Church's collectives is a consequence of the general result of Abraham  Wald \cite{[173]} which we formulate now.

Let $p=(p_j=P(\alpha_j))$ be a probability
distribution on the label set $L=\{\alpha_1,..., \alpha_m\}.$ 
Denote the set of all possible sequences with elements from $L$ by the symbol $L^\infty.$ Let $\phi$ 
be a place selection. For $x \in L^\infty,$ the symbol $\phi x$ is used to denote the subsequence 
of $x$ obtained with the aid of this place selection.
  
Let $U$ be some family of place selections. We set
$$
X(U; p)= \{x \in L^\infty: \forall \phi \in U \; \lim_{N \to \infty} \nu_n(\alpha_j; \phi x)=p_j, j=1,...,m\},
$$
where as usual $\nu_N(\alpha;y), \alpha \in L,$ denotes the relative frequency of the appearance of the label $\alpha$ 
among the first $N$ elements of the sequence $y \in L^\infty.$ 

\medskip

{\bf Theorem 2.}  (Wald). {\it For any countable set $U$ of place selections and any probability
distribution $p$ on the label set $L,$ the set of sequences $X(U; p)$ has the cardinality of
the continuum.}

\medskip

By Wald's theorem for any countable set of place selections $U$ the frequency theory
of probability can be developed at the mathematical level of rigorousness. R. von
Mises was completely satisfied by this situation (see \cite{[171]}). 

However, a new cloud appeared on the sky. This was the famous  
{\it Ville's objection \cite{Ville}.}

{\bf Theorem 3.} (Ville). {\it Let $L=\{0,1\}$ and let $U=\{\phi_n\}$ be a countable set
of place selections. Then there exists $x\in  L^\infty$  such that
\begin{itemize}

\item  for all $n,$ 
$$
\lim_{N\to \infty} \sum_{j=1}^N (\phi_n x)_j= 1/2;
$$

\item for all $N,$ 
$$
\sum_{j=1}^N (\phi_n x)_j \geq 1/2.
$$ 
\end{itemize}
}

Such a sequence $x$ of zeros and ones is a collective with respect to $U,$ $x \in X(U;1/2),$ but seems to be far too
regular to be called random.

At the same time from the measure-theoretic viewpoint  the existence of such sequences is not a problem. The set of such sequences 
has the Lebesgue measure zero.

We recall that any sequence of zeros and ones can be identified with a real number from the segment 
$[0,1]:$
$$
x=(x_1,...,x_N,...) \to r= x_1/2+ ...+ x_N/p^{N+1}+... \in \mathbf{R}
$$
 On this segment we have the linear Lebesgue measure which is defined through its values on intervals: $\mu([a, b))= b-a.$ 
And the previous statement is about this measure.

Here we see the difference between the treatment of randomness as unpredictability (a la von Mises) and as typicality (see section 
\ref{MLOF} for the latter -- theory of Martin L\"of).
   
\subsection{Approach of Per Martin L\"of: randomness as typicality} 

\label{MLOF}

Ville  used Theorem 3 to argue that collectives in the sense of von Mises
and Wald do not necessarily satisfy all intuitively required properties of randomness.

Jean Ville introduced  \cite{Ville} a new way of characterizing random sequences, based on the following
principle:

\medskip

{\bf Ville's Principle:} {\it A  random sequence should satisfy all properties of probability one.}

\medskip

Each property is considered as determining a {\it  test of randomness.} According to Ville \cite{Ville}, a sequence 
can be considered as a random if it passes all possible tests for randomness.

However, this is impossible: we have to choose countably many from among
those properties (otherwise the intersection of the uncountable family of set of probability 1 can have probability less than 1 or simply be 
nonmeasurable; in the latter case the probability is not defined at all). Countable families of properties (tests for randomness) 
can be selected in various ways. A random sequence passing one countable sequence of tests can be rejected by another. This brings ambiguity in the 
 Ville
approach to randomness as typicality (i.e., holding some property with probability one).

It must be underlined that the Ville principle is really completely foreign to
von Mises. For von Mises, a collective $x\in  L^\infty$  induces a probability on the set of labels
$L,$ not on the set of all sequences $L^\infty.$ Hence, for von Mises (and other scientists interpreting randomness as unpredictability 
in an individual sequence),  there is no connection  between
properties of probability one in $L^\infty$ and properties of an individual collective.

Later (in 1970th) Per Martin-L\"of (motivated by the talk of Andrei Nikolaevich Kolmogorov at 
the Moscow Probability Seminar) \cite{[142]} 
solved the problem of ambiguity of the Ville interpretation of randomness as typicality. He
proposed to consider {\it recursive properties of probability
one}, i.e., the properties which can be tested with the aid of algorithms. Such an
approach induces the fruitful theory of recursive tests for randomness (see, for example,
\cite{[143]}, \cite{[144]}). The key point of this ``algorithmic tests'' approach to the notion of randomness is that it is possible 
to prove that there exists the {\it universal algorithmic test}.  A sequence is considered as random if it passes this 
universal test. Thus the class of typicality-random sequences is well defined. Unfortunately, this universal test of randomness
cannot be constructed algorithmically (although by itself it is an algorithmic process). Therefore, although we have the well 
defined class of Martin L\"of random sequences, we do not know how the universal test for such randomness looks. Hence, for the concrete 
sequence of trials we cannot check algorithmically whether it is random or not -- although we know that it is possible to perform
such algorithmic check. 

So, as in the case of von Mises/Wald randomness (as unpredictability),  Ville/Martin L\"of  randomness (as typicality) 
cannot be considered as satisfactory.  

\medskip

Finally, we remark that similar approach, randomness as (recursive) typicality,
 was developed by Schnorr \cite{[158]}.\footnote{We state again
that approaches of Martin-L¨of and Schnorr (as well as Ville) have nothing
to do with the justification of Mises' frequency probability theory and his viewpoint on 
randomness as unpredictability.}

\subsection{Approach of Andrei Nikolaevich Kolmogorov: randomness as complexity}

It is well know that personally A. N. Kolmogorov was not satisfied by his own measure-theoretic approach 
to probability (private communications of his former students). He sympathized to the von Mises approach 
to probability for which randomness was not less fundamental than probability. In 60th he turned again to the 
foundations of probability and randomness and tried to find foundations of randomness by reducing this notion to
the notion of complexity.  Thus in short the Kolmogorov approach can be characterized as {\it randomness as complexity.}
Another basic point of his approach is that complexity of a sequence has to be checked {\it algorithmically.}

Let $L=\{0,1\}.$ Denote by $L^*$ the set of all {\it finite sequences} (words)  in the alphabet $L.$ 

\medskip

{\bf Definition 8.1} (Kolmogorov). {\it Let $A$ be an arbitrary algorithm. The complexity of a
word $x$ with respect to $A$ is $K_A(x)=  \min l(\pi),$ where $\{\pi\}$ are the programs which
are able to realize the word $x$ with the aid of $A.$}

\medskip

Here $l(\pi)$ denotes the length of a program $\pi.$ This definition depends on the structure
of an algorithm $A.$ Later Kolmogorov proved the following fundamental theorem:

\medskip

{\bf Theorem 4.} {\it There exists an algorithm $A_0$ (optimal algorithm) such that, for any 
algorithm $A,$ there exists a constant} $C>0,$
\begin{equation}
\label{HH}
K_{A_0}(x) \leq  K_{A}(x) +C .
\end{equation}

\medskip

It has to be pointed out that optimal algorithm is not unique. 

\medskip

The complexity $K(x)$  of the word $x$ is (by definition) equal to the complexity $K_{A_0}(x)$ 
with respect to one fixed (for all considerations) optimal algorithm $A_0.$

The original idea of Kolmogorov  \cite{[134]a, [134], [135]} was that complexity $K(x_{1:n})$  of the initial
segments $x_{1:n}$  of a random sequence $x$ has to have the asymptotic 
$\sim n$
\begin{equation}
\label{HHa}
K(x_{1:n}) \sim n , n \to \infty,
\end{equation}
i.e., we might not find a short code for $x_{1:n}.$

However, this heuristically attractive idea idea was rejected as a consequence of  the  
objection of Per Martin-L\"of \cite{[144]}.

To discuss this objection and connection of the Kolmogorov complexity-randomness 
with Martin-L\"of typicality-randomness, we proceed with
{\it conditional Kolmogorov complexity} $K(x; n),$ instead of complexity $K(x).$ Conditional complexity
$K(x; n)$ is defined as the length of the minimal program $\pi$ which produces the output
$x$ on the basis of information that the length of the output $x$ is equal to $n.$  

\medskip

{\bf Theorem 5.} (Martin L\"of) {\it   For every binary sequence $x,$
\begin{equation}
\label{HHb}
K(x_{1:n};n) < n -\log_2 n,
\end{equation}
for infinitely many n.}

\medskip

Hence, {\it Kolmogorov random sequences, in the sense of the definition (\ref{HHa}),
do not exit.}

Another problem of the Kolmogorov approach to randomness as algorithmic complexity is that we ``do not know'' any
optimal algorithm $A_0,$ i.e., the Kolmogorov complexity is not algorithmically computable. However, the latter 
problem can be partially fixed, because there exist algorithmic methods to estimate this complexity (from above and below).

We remark that historically the Kolmogorov algorithmic complexity approach to the notion of randomness preceded the Martin L\"of 
recursive test approach and Per Martin L\"of as staying in Moscow at that time was influenced by the ideas of Kolmogorov.

Finally, we remark that {\it numerical evaluation of the Kolmogorov complexity for binary sequences produced in experiments with quantum systems
is an open and interesting problem.}

\subsection{Randomness: concluding remarks}
\label{FINAL}

As we have seen, non of the three basic mathematical approaches to the notion of randomness (based on unpredictability, typicality, and algorithmic 
complexity) led to the consistent and commonly accepted theory of randomness. Can we hope to create such a mathematical theory? 
As was already remarked, Kolmogorov died with the hope that in future  a new and unexpected approach to the notion of randomness 
will be elaborated. And, of course, mathematicians will continue to work on this problem. However, it may happen that  future 
attempts will never lead to a mathematically acceptable notion of randomness. This was the final point of my first lecture given 
at IQOQI. 

In the after-talk discussion various opinion were presented; in particular, prof. Zeilinger conjectured that such a painful
process of elaboration of the mathematical theory of randomness is simply a consequence of the methodological mistreatment of this notion.
It might be that randomness is not mathematical, but physical notion. Thus one has to elaborate physical procedures guarantying randomness
of experimentally produced sequences and not simply try to construct such procedures mathematically. In some sense Zeilinger's proposal 
is consonant with von Mises' proposal: to find a collective, one simply has to go to casino.

\newpage

\section{Appendix: Around the Kolmogorov measure-theoretic  and von Mises frequency axiomatics}
\label{APP}

\subsection{On generalizations of the measure-theoretic probability theory}

Kolmogorov by himself did not consider his axiomatics as something final and unchangeable. He discussed various possible modifications
of the axiomatics. In particular, for him the request that the collection of all possible events has to form a $\sigma$-algebra or even 
an algebra of sets was questionable. He proposed some models of ``generalized probability'' where the collection of events need not have 
the standard structure of ($\sigma$-)algebra. 

However, later the modern axiomatics of probability  was so to say ``crystallized'' and any 
departure from the Kolmogorov axiomatics was considered as a kind of pathology. 

Nevertheless, the concrete applications continuously 
generate novel models in which probabilities have some unusual features. For example, in psychology and game theory 
non-additive ``probabilities'' were invented and actively used, see, e.g., \cite{UB_KHR} for references. We also mention ``negative probabilities''
which were explored in the works of Dirac, Feynman, Aspect  
and my own works, see \cite{AspectN}, \cite{FeynmanN}, \cite{KHR_INT}; 
see also  M\"uckenheim \cite{MUK} for the detailed review on negative probabilities 
in quantum physics. Dirac even used 
``complex probabilities'' \cite{DiracC}.

\subsection{Negative probabilities}

``Negative probabilities'' appear with strange regularity in majority of problems of quantum theory starting with Dirac relativistic quantization 
of the electromagnetic field. Feynman was sure that all quantum processes can written with the aid of negative transition probabilities which disappear
in the final answers corresponding to the results of measurements. Of course,  the interpretation of signed measures taking, in particular,
 negative values as probabilities (in particular, the use of signed density functions)  
are counterintuitive. Nevertheless, operation with such generalized probabilities is less counterintuitive than operation in the formal quantum framework. 
One can say (as Feynman often did) that the only difference between classical and quantum physics is that in the latter one has to use 
generalization of the Kolmogorov measure-theoretic model with signed measures, instead of positive measures.

The mathematical formalism of theory of negative probabilities was presented in my book \cite{KHR_INT0} (and its completed edition
\cite{KHR_INT}). Mathematically this is a trivial generalization of the Kolmogorov notion of probability space.

Let again  $\Omega$ be a set and let $\mathcal{F}$ be a $\sigma$-algebra of its
subsets. A \textit{signed probability measure} $P $ is a map from
$\mathcal{F}$ to the segment $\mathbf{R}$ normalized $P(
\Omega)=1$ and $\sigma$-additive. The \textit{signed probability space} \cite{KHR_INT0}
\index{signed probability space} is a triple $\mathcal{P}=$
$(\Omega,$ $\mathcal{F},$ $P).$ Points $\omega$ of $\Omega$ are said to be \textit{elementary events},
elements of $\mathcal{F}$ are \textit{random events}, $P$ is
\textit{signed probability}.%
\index{signed probability}%

If such  a ``probability'' takes negative values, then automatically it has to take values exceeding 1. It is easy to show. 
Take an event $A$ such that $P(A)<0.$ Take its complement $B= \Omega \setminus A.$ Then additivity and normalization by 1 imply that $P(\Omega)= 
P(A\cup B)= P(A) + P(B)=1,$ i.e., $P(B) >1.$ 

On the basis of the signed probability space it is possible to proceed quite far by generalizing the basic notions and constructions 
of the standard probability theory: conditional probabilities are defined by the Bayes formula, independence is defined via factorizing 
of probability, the definition of a random variable is the same as in the standard approach, average (mean value) is given by the 
integral.\footnote{We remark that each signed measure can be canonically  represented as the difference of two positive measures
(the Jordan decomposition of a signed measure). Thus one can define
the integral with respect to a signed measure as the difference of integrals with respect to these two positive measures.}  

The complex probability space is defined in the same way; here ``probabilities'' are normalized $\sigma$-additive functions taking values in 
the field of complex numbers $\mathbf{C}.$ 

The prejudice  against such generalized probabilities is the result of a few centuries of the  use of the Laplacian definition of probability as the proportion
$P(A)=m/n,$ where $m$ is the number of cases favoring to the event $A$ and the $m$ is the total number of possible cases. Take, for example,
a coin,  then here $n=2$ and, for the event $H$ that the head will appear in one of trials, $m=1.$ Hence, $P(H)=1/2.$ For the dice, we
shall get probabilities $P(A_{j})=1/6,$ where the index $j=1,2,..,6$ labels the corresponding sides of the dice.  It is clear that the Laplacian 
probability cannot be negative or exceed 1. However, the domain of application of the  Laplacian 
probability is very restricted. For example, it can be used only for symmetric coins and dices. (Nevertheless, since  the Laplacian 
probability is the starting point of all traditional courses in probability theory, the notion of probability is often associated 
precisely with its  Laplacian version. In particular, this image of probability was strongly criticized by Richard von Mises \cite{[170], [171]}.)    
If one assign weights to events by using not only the Laplacian rule (which is applicable only in very special cases), then, in principle,
one may try to use any real number as the weight of an event. There is the order structure on the real line. Thus it is possible 
to compare probabilities and to say that one event is more probable than another.
Hence, the (b)-part of the Kolmogorov interpretation of probability, see section \ref{INTK},
is also applicable to signed probabilities. 

For signed probabilities, the possibility to use the frequency interpretation of probability, the (a)-part of the Kolmogorov interpretation (see section \ref{INTK}), is a 
more delicate question. Surprisingly there were obtained analogs of the law of large numbers and the central limit theorem.\footnote{
It is convenient to proceed with complex ``probabilities'', i.e., by considering the signed probabilities as a particular case of 
complex probabilities, see \cite{SUPER} for the most general case: the limit theorems were derived for ``probabilities'' taking values 
in complex Banach (super)algebras. This possibility to generalize the basic limit theorems of the standard probability theory to real and complex 
valued probability measures (as well as to more general measures) is also a supporting argument to use, e.g., negative probabilities. Sometimes
mathematics can lead us to areas where our intuition deos not work.} However, it is impossible to obtain such a strong type of 
convergence as we have for 
the standard positive probabilities. For example, in the generalized law of large numbers the  arithmetic mean need not converge to the probabilistic
mean $m$ almost everywhere. For identically distributed and independent random variables $\xi_1,...,\xi_N,...,$ with the mean value 
$m=E \xi_j= \int_{\Omega} \xi_j(\omega) dP(\omega),$  it can happen that 
\begin{equation}
\label{NEG}
P\Big(\omega\in \Omega: \frac{\xi_1(\omega) +...+\xi_N(\omega)}{N} \to m, N \to \infty \Big)\not=1.
\end{equation}
Nevertheless, a weaker convergence is still possible. Set 
$$
\eta_N= \frac{\xi_1(\omega) +...+\xi_N(\omega)}{N}.
$$
Then it is possible to show that, for a sufficiently extended class of functions, $f: \mathbf{R}  \to  \mathbf{R},$ 
the following form of the weak convergence takes place:
\begin{equation}
\label{NEG1}
E f(\eta_N) = \int_\Omega  f(\eta_N(\omega)) dP(\omega) \to f(m), N \to \infty.
\end{equation}  
We remark that, for the standard positive probability,  the limit relation (\ref{NEG1}) holds for all bounded continuous functions
(this is a consequence of the strong law of large numbers). 
For generalized, signed or complex, 
probabilities   (or even with values in Banach algebras) functions for which (\ref{NEG1}) holds have to have 
some degree of smoothness \cite{SUPER}.

\subsection{On generalizations of the frequency theory of probability}

Another way to obtain the frequency interpretation of negative probabilities is to use the number-theoretic approach based on so called
$p$-adic numbers. The starting point is that frequencies $\nu_n=n/N$ are always {\it rational numbers.} We recall that the set of rational numbers 
$\mathbf{Q}$ is a dense subset in the set of real numbers $\mathbf{R},$ i.e., the real numbers can be obtained as limits of rational 
numbers. Von Mises implicitly  explored this number-theoretic fact in the formulation of the principle of the statistical stabilization 
of relative frequencies. For example, the probability of an event is a real number (in $[0,1]),$ because he considered one special 
convergence on $\mathbf{Q},$ namely, with respect to the metric induced from $\mathbf{R}: \rho_{\mathbf{R}}(x,y)= \vert x - y\vert.$ 
The additivity of the frequency probability is the result of the continuity of the operation of addition with respect to this metric.
It can shown \cite{KHR_INT} that to get the Bayes formula, one has to use the continuity of the operation of division
in $\mathbf{R}\setminus \{0\}.$
Thus the von Mises theory of frequency probability is based on the dense embedding 
\begin{equation}
\label{EBD}
\mathbf{Q} \subset \mathbf{R}.
\end{equation}
 It is important that $\mathbf{R}$ is a {\it topological field}, i.e., all algebraic operations on it are continuous.

This analysis of the number-theoretic-topological structure of the von Mises model leads us to the following  natural question:

{\it Is it possible to find dense embedding of the set of rational numbers $\mathbf{Q}$ (representing all relative frequencies) 
into another topological number field $K,$ $\mathbf{Q} \subset K$?}

If the answer were positive, then one would be able to extend the domain of application of 
 the principle of the statistical stabilization of von Mises: to consider the problem of the existence of the
limits of relative frequencies in the topology of the number field $K.$ Since the standard properties of probability were based on the consistency 
between the algebraic and  topological structures, one can expect that, for a topological field, the corresponding generalization 
of the (frequency) probability theory will be very similar to the standard one based on the embedding (\ref{EBD}).

\subsection{$p$-adic probability}

It is interesting that there are not so many ways to construct dense embeddings of $\mathbf{Q} \subset \mathbf{K}$ 
(in other words: to complete the field of rational numbers with respect to some metric and to obtain a topological field).
By the famous theorem of number theory, {\it the Ostrowsky theorem}, all ``natural embeddings'' are reduced  to the fields of so called {\it $p$-adic numbers} $\mathbf{Q}_p,$
where $p=2,3,..., 1997, 1999,...$ are prime numbers. The elements of these fields have the form:
$$
a=\sum_{j=k}^\infty a_j p^j, a_j=0,1,.., p-1, 
$$
where $k$ is an integer; in particular, for $p=2$ this a binary expansion. We recall that each real number can be represented as 
$$
a=\sum_{- \infty}^k a_j p^j, a_j=0,1,.., p-1. 
$$
Thus a real number can contain infinitely many terms with the negative powers of $p$ and a $p$-adic number can contain infinitely many terms with 
the positive powers of $p.$  

In principle, for some sequences $x=(x_1,...,x_n,...)$ of the results of trials (where $x_j\in L$ and $L$ is the corresponding label set),
the limits of relative frequencies may exist not in $\mathbf{R},$ but in one of $\mathbf{Q}_p.$ Such a ``collective'' wil generate 
probabilities belonging not to $\mathbf{R},$ but to $\mathbf{Q}_p.$ One have not to overestimate the exoticness of such ``probabilities''.
One has to understand well that the real numbers (opposite to the rational numbers) are just symbols (representing limits of sequence 
of rational numbers); the $p$-adic numbers are similar symbols. 

Foundations of the $p$-adic probability theory can be found in \cite{KHR_INT}. This theory was developed in the framework of 
$p$-adic theoretcial.One of the interesting mathematical facts about such generalized 
probabilities is that the range of the values of $p$-adic probabilities coincides with $\mathbf{Q}_p,$ i.e., any $p$-adic number can be 
in principle obtained as the limit of relative frequencies, $a= \lim_{N \to \infty} n/N$ (cf. with the standard frequency probability theory 
which implies that the range of values of probability coincides with the segment $[0,1]$ of the real line). In particular, any rational number
(including negative numbers) can be obtained in this way.

Therefore in the $p$-adic framework ``negative probabilities'' appear in the natural way.
Typically convergence of a sequence of rational numbers (in particular, relative frequencies) in $\mathbf{Q}_p$ implies that this sequence
does not converge in $\mathbf{R}$ and vice versa, see \cite{PADIC} for examples. In this framework the appearance of the $p$-adic and, in particular, negative
probabilities, can be interpreted as a sign of irregular (from the viewpoint of the standard probability theory) 
statistical behavior, as a violation
of the basic principle of the standard probability theory: the principle of the statistical stabilization of relative frequencies (with respect 
to the real metric on $\mathbf{Q}).$ In the measure-theoretic approach this situation can be considered as a violation of the law of large numbers. 

In short: negative probability means that, for some event $A,$ in the standard  decimal (or binary) expansion of the relative frequencies 
$\nu_N(A)= n(A)/N$ the digits would not stabilize when $N \to \infty.$  We remark that, in principle, there is nothing mystical in such behavior
of data, see again for examples \cite{PADIC}.

\subsection{Violation of the principle of the statistical stabilization for hidden variables?}

 As was pointed out, for Feynman and Aspect hidden variables have to be described by signed 
probability distributions. In the light of the previous considerations this statement can be interpreted in the very natural manner: 
relative frequencies for hidden variables, $\nu_N(\lambda)= n_N(\lambda)/N,$  do not satisfy the principle of the statistical stabilization. In this way the realism can be saved:
the local hidden variables are not forbidden, but the principles of the standard probability theory are not applicable to them. 
However, it is not clear whether such a viewpoint leads to reestablishing of realism at the subquantum level. My personal impression is that 
realism by itself is nothing else than the acceptance of the global validity of the principle of the statistical stabilization (the law of large 
numbers). In seems that all our senses were formed to recognize only statistically repeatable phenomena (and what is important: statistically repeatable
with respect to the real metric). As a consequence all our measurement devices are designed to observe only such 
phenomena.\footnote{To be honest, I was not able to find experimental statistical data violating the standard principle of the  statistical stabilization.
Although in the book \cite{PADIC} the reader can find various examples of stochastic evolution of biological systems in which statistical data 
violate the principle of the statistical stabilization, these data is so to say about hidden features of such biological evolutions.} 

What is still unclear for me is how the continuous space-time structure (modeled with the aid of real numbers) is transferred  into the
statistical stabilization of frequencies of observed events. It is clear that all our measurement devices explore this space-time structure.
Are they just machines which transfer features of the continuous space-time into relative frequencies stabilizing with respect to the real metric
and leading to the standard models of probability? or another way around? Do we get the standard model of space-time just because our senses
are based on the principle of the statistical stabilization with respect to the real metric? In fact, the convergence with respect to the real metric
is based on the agreement  that for large $N$ its inverse $1/N$ is a small quantity, if $N$ is very large than $1/N$ is negligibly small and
it can be neglected. Is this the essence of modern probability theory and even the modern science in general?


\begin{thebibliography}{99}

\bibitem{Boole} Boole, G.: An Investigation of the Laws of Thought. Dover Edition, New York (1958)

\bibitem{Boole1} Boole, G.: On the theory of probabilities. Phil. Trans. Royal Soc. London, 152, 225-242 (1862)


 \bibitem{BR1}    Brukner, C.,  and Zeilinger, A.,  Malus' law and quantum information. Acta Physica Slovava, 49(4):647-652, 1999.

 \bibitem{BR2}    Brukner, C.,  and Zeilinger, A., Operationally invariant information in quantum mechanics. Physical Review Letters, 83(17):3354-3357, 1999.
 
\bibitem{BR3} Brukner, C.,  and Zeilinger, A., Information Invariance and Quantum Probabilities, Found. Phys. 39, 677 (2009).

\bibitem{Chaitin} Chaitin, G. J.,  On the Simplicity and Speed of Programs for Computing Infinite Sets of Natural Numbers. 
Journal of the ACM 16 (1969), 407–422.

\bibitem{D1} Chiribella, G.,  D'Ariano, G. M.,  and Perinotti, P.,  
Informational derivation of quantum theory.
Phys. Rev. A 84, 012311  (2011).

\bibitem{Fuchs} Fuchs, C. A.: The anti-V\"axj\"o interpretation of quantum mechanics.
In: Khrennikov, A.Yu. (ed) Quantum Theory: Reconsideration of
Foundations,  pp. 99-116. Ser. Math. Model. 2, V\"axj\"o
University Press, V\"axj\"o  (2002)

\bibitem{Fuchs1} Fuchs, C. A.: Quantum mechanics as quantum information (and only a little more).
In: Khrennikov, A.Yu. (ed) Quantum Theory: Reconsideration of
Foundations,  pp. 463-543. Ser. Math. Model. 2, V\"axj\"o
University Press, V\"axj\"o  (2002)

\bibitem{Fuchs2} Fuchs, C.: Delirium quantum (or, where I will take quantum mechanics if it will let me). In:
Adenier, G., Fuchs, C. and Khrennikov, A. Yu. (eds) Foundations of
Probability and Physics-3, pp. 438-462.  American Institute of
Physics, Ser. Conference Proceedings 889, Melville, NY (2007)


\bibitem{Beyond} Khrennikov, A., Beyond Quantum.
Pan Stanford Publishing; Singapore, 2014.  

\bibitem{KHR_INT0} Khrennikov, A., Interpretations of Probability. VSP Int.
Sc. Publishers, Utrecht/Tokyo, 1999.

\bibitem{KHR_INT} Khrennikov, A.,  Interpretations of Probability. De Gruyter,
Berlin, 2009,  second edition (completed).

\bibitem{Vaxjo1}  Khrennikov, A.,
V\"axj\"o Interpretation of Quantum Mechanics. Published: Proc. Int. Conf. 
Quantum Theory: Reconsideration of Foundations. 
Ser. Math. Modelling in Physics and Cognitive Sciences, 
V\"axj\"o Univ. Press, volume 2, p.163-170, 2002; 
arXiv:quant-ph/0202107

\bibitem{Vaxjo}  Khrennikov, A.,  V\"axj\"o interpretation-2003: Realism of
contexts. Proc. Int. Conf. {\it Quantum Theory: Reconsideration of
Foundations.} Ser. Math. Modelling in Phys., Engin., and Cogn.
Sc., {\bf 10}, 323-338, V\"axj\"o Univ. Press, 2004.

\bibitem{UB_KHR} Khrennikov, A.,  Ubiquitous  quantum structure: from
psychology to finances, Springer, Berlin-Heidelberg-New York,
2010.

\bibitem{KHR_ARXIV} Khrennikov, A., CHSH inequality: Quantum probabilities as classical conditional probabilities.
	arXiv:1406.4886 [quant-ph].

\bibitem{K} Kolmolgorof A. N., Grundbegriffe der Wahrscheinlichkeitsrechnung, Springer-Verlag,
Berlin, 1933.

\bibitem{K1} Kolmolgorov A. N., Foundations of the Probability Theory, Chelsea Publ. Comp.,
New York, 1956.

\bibitem{K2} Kolmogorov A. N. The bsic notions of probability theory. Nauka, Moscow, 1974 (in Russian).

\bibitem{[134]a}  Kolmogorov, A. N.,  On Tables of Random Numbers. Sankhya Ser. A. 25  (1963) 369–375.

\bibitem{[134]} Kolmolgorov A. N., Three approaches to the quantitative definition of information,
Problems Inform. Transmition. 1 (1965), 1–7.

\bibitem{[135]} Kolmolgorov A. N., Logical basis for information theory and probability theory, IEEE
Trans. IT–14 (1968), 662–664.
   
\bibitem{[142]} Martin-L\"of P., On the concept of random sequence, Theory of Probability Appl. 11
(1966), 177–179.

\bibitem{[143]} Martin-L\"of P., The definition of random sequence, Inform. Contr. 9 (1966), 602–619.

\bibitem{[144]} Martin-L\"of P., Complexity oscillations in infinite binary sequences, Z. Wahrscheinlichkeitstheorie
verw. 19 (1971), 225–230.

\bibitem{[158]} Schnorr C. P., Zuf¨alligkeit und Wahrscheinlichkeit, Lect. Notes in Math. 218, Springer-
Verlag, Berlin, 1971.

\bibitem{Ville} Ville J., Etude critique de la notion de collective, Gauthier-Villars, Paris, 1939.

\bibitem{[169]} von Mises R., Grundlagen der Wahrscheinlichkeitsrechnung, Math. Z. 5 (1919), 52–99.

\bibitem{[170]} von Mises R., Probability, Statistics and Truth, Macmillan, London, 1957.

\bibitem{[171]} von Mises R., The mathematical theory of probability and statistics, Academic, London,
1964.

\bibitem{VN} von Neuman J., Mathematical foundations of quantum mechanics, Princeton Univ.
Press, Princenton, 1955.

\bibitem{V} Vorob'ev, N. N., Consistent families of measures and their extensions. 
Theory of Probability and its Applications 7,  147-162 (1962)

\bibitem{[173]} Wald A., Die Widerspruchsfreiheit des Kollektivbegriffs in der Wahrscheinlichkeitsrechnung,
Ergebnisse eines Math. Kolloquiums 8, 38–72 (1938).

\bibitem{Z0} Zeilinger, A., A foundational principle for quantum mechanics. Foundations of Physics 29 (1999), 631- 641.

\bibitem{Z1} Zeilinger, A.,   Dance of the Photons: From Einstein to Quantum
Teleportation. Farrar, Straus and Giroux, New-York, (2010)


\end{thebibliography}

\begin{thebibliography}{99}

\bibitem{AspectN} Aspect, A., Comment on ``A classical model of the EPR experiment with quantum mechanical correlations and Bell 
inequalities''. In: Frontiers of Nonequilbrium Statistical Physics, Moore, G. T., and Sculy, M. O. (eds). Plenum Press, New York, London,
pp. 185-189.

\bibitem{DiracC}  Dirac, P. A. M., The physical interpretation of quantum mechanics. Proc. Roy. Soc. London
A180 (1942), 1-39.

\bibitem{FeynmanN}  Feynman, R. P., Negative probability, in: Quantum Implications, Essays in Honour of
David Bohm, edited by B. J. Hiley and F. D. Peat, pp. 235-246, Routledge and Kegan
Paul, London, 1987.

\bibitem{SUPER} Khrennikov, A.,  {\it Supernalysis.} Nauka, Fizmatlit, Moscow,
1997 (in Russian). English translation: Kluwer, Dordreht, 1999.

\bibitem{PADIC} Khrennikov, A., {\it Non-Archimedean analysis: quantum
paradoxes, dynamical systems and biological models.} Kluwer,
Dordreht, 1997.

\bibitem{MUK} M\"uckenheim, W., A review on extended probabilities, Phys. Reports 133 (1986), 338–
401.


  
\end{thebibliography}
\end{document}